\documentclass[conference]{IEEEtran}
\IEEEoverridecommandlockouts

\usepackage{marvosym}
\usepackage{authblk}
\usepackage{setspace}
\usepackage{cite}
\ifCLASSINFOpdf
	\usepackage[pdftex]{graphicx}
\else
	\usepackage[dvips]{graphicx}
\fi
\usepackage{amsmath,amssymb,amsthm}
\usepackage{caption}
\usepackage{amsfonts}
\usepackage{algorithm,algorithmic}
\usepackage{array}
\usepackage{makecell}				
\usepackage{cite}
\usepackage{esint}
\usepackage{enumerate}
\usepackage{times}
\usepackage{url}
\usepackage{stfloats}
\usepackage{subfigure}

\ifCLASSOPTIONcompsoc
  \usepackage[caption=false,font=normalsize,labelfont=sf,textfont=sf]{subfig}
\else
  \usepackage[caption=false,font=footnotesize]{subfig}
\fi

\newcommand{\PreserveBackslash}[1]{\let\temp=\\#1\let\\=\temp}
\newcolumntype{C}[1]{>{\PreserveBackslash\centering}p{#1}}
\newcolumntype{R}[1]{>{\PreserveBackslash\raggedleft}p{#1}}
\newcolumntype{L}[1]{>{\PreserveBackslash\raggedright}p{#1}}

\begin{document}

\title{Joint Optimization of Active and Passive Beamforming in Multi-IRS Aided mmWave Communications\\
	\thanks{This work was supported in part by the NSFC under Grant 62001071, 62101460 and 62061007, the Macao Young Scholars Program under Grant AM2021018, the China Postdoctoral Science Foundation under Grant 2020M683291, and the Science and Technology Research Program of Chongqing Municipal Education Commission under Grant KJQN202200617. The work of S. Ma was supported in part by the Science and Technology Development Fund, Macau SAR under Grants 0051/2022/A1, 0036/2019/A1 and SKL-IOTSC (UM)-2021-2023, and in part by the Research Committee of University of Macau under Grant MYRG2020-00095-FST. The work of Y. Xu was supported in part by the NSFC under Grant 62271094 and the Scientific and Technological Research Key Program of Chongqing Municipal Education Commission under Grant KJZD-K202200601. (\textit{Corresponding author: Qing Xue.})}
}

\author[1]{Renlong~Wei}
\author[1,2]{Qing~Xue}
\author[2,3]{Shaodan~Ma}
\author[1]{Yongjun~Xu}
\author[4]{Li~Yan}
\author[4]{Xuming~Fang}

\affil[1]{\small \textit{School of Communication and Information Engineering, Chongqing University of Posts and Telecommunications, Chongqing 400065, China}}
\affil[2]{\textit{State Key Laboratory of Internet of Things for Smart City, University of Macau, Macao SAR, China}}
\affil[3]{\textit{Department of Electrical and Computer Engineering, University of Macau, Macao SAR, China}}
\affil[4]{\textit{Key Lab of Information Coding \& Transmission, Southwest Jiaotong University, Chengdu 610031, China}
	\authorcr E-mails: s210101140@stu.cqupt.edu.cn, xueq@cqupt.edu.cn,
	\authorcr shaodanma@um.edu.mo, xuyj@cqupt.edu.cn, liyan@swjtu.edu.cn, xmfang@swjtu.edu.cn}

\maketitle

\begin{abstract}
Intelligent reflecting surface (IRS) has been considered as a promising technology to alleviate the blockage effect and enhance coverage in millimeter wave (mmWave) communication. To explore the impact of IRS on the performance of mmWave communication, we investigate a multi-IRS assisted mmWave communication network and formulate a sum rate maximization problem by jointly optimizing the active and passive beamforming and the set of IRSs for assistance. The optimization problem is intractable due to the lack of convexity of the objective function and the binary nature of the IRS selection variables. To tackle the complex non-convex problem, an alternating iterative approach is proposed. In particular, utilizing the fractional programming method to optimize the active and passive beamforming and the optimization of IRS selection is solved by enumerating. Simulation results demonstrate the performance gain of our proposed approach.
\end{abstract}


\IEEEpeerreviewmaketitle

\section{Introduction}
As one of the key technologies in 5G and beyond systems, millimeter wave (mmWave) has huge spectrum resource and can support extremely high data rates \cite{MIMO-OTFS} However, mmWave signals suffer high propagation path loss and are susceptible to blockage due to the short wavelength. To compensate the high path loss, massive antenna array technology is widely adopted, but it fails to mitigate the blockage problem \cite{mmWave-MIMO}. Using ultra-dense network is an effective way to alleviate blockages, but it involves more serious interference and power consumption than traditional networking \cite{UDN}, \cite{UDN-2}.

Recently, intelligent reflecting surface (IRS), which is composed of many low-cost reflecting elements, has attracted much attention. By altering the phase shifts or/and amplitudes, each element of IRS can reflect incident signal to receiver independently \cite{IRS-6G}, \cite{LISA}. Deploying IRS in wireless communication system can intelligently reconfigure wireless environment and increase the received signal strength by controlling the direction of the reflected signal. This provides an alternative approach to tackle the problems in mmWave system with low-cost and energy-efﬁcient.

Many studies on IRS aided wireless communications \cite{IRS-Cell-free-network,IRS-active-passive,IRS-MIMO-up,CSIT,Sum-rate-1,Sum-rate-3} have been conducted since IRS has numerous potential benefits. In the IRS-aided wireless systems, the effective beamforming is the key to achieve the goal of system performance optimization. However, compared with traditional beamforming, the beamforming problem in IRS-aided system is more difficult to solve because it involves two parts: the active beamforming at base station (BS) and the passive beamforming at IRS. The performance of the two parts is closely related to each other. Many efforts have been dedicated to the research of the beamforming in IRS assisted system. For example, the authors of \cite{IRS-active-passive} investigated the transmit power minimization by jointly optimizing the active and passive beamforming. In \cite{Sum-rate-1}, the weighted sum-rate maximization problem is studied and the active and passive beamforming are solved by quadratic transform method and quadratically constrained quadratic program (QCQP) respectively. However, only one IRS is considered in most of these works. Due to the  severe path loss and the unique channel characteristics of mmWave communications, it is very necessary to deploy multiple IRSs discretely to ensure the performance of mmWave systems. Meanwhile, to improve the system coverage, multiple IRSs are usually deployed far away from each other. In this case, the assistance effect of lRSs deployed in different locations usually varies significantly. Hence, considering the optimal IRS selection for different user is necessary \cite{IRS-U}. In this work, when deal with the beamforming optimization problem, we jointly consider the optimal IRS selection in the multi-IRS aided mmWave system. After introducing multi-IRS and the IRS selection in the mmWave system, the joint beamforming becomes more complex and challenging, so an effective joint optimization method needs to be redesigned.

Motivated by the above, to guarantee the robustness of the multi-user downlink transmissions, we consider a multi-IRS assisted mmWave communication system and mainly focus on the transmit beamforming at mmWave BS (mBS) and passive beamforming at the IRS and jointly study the IRS selection problem. In order to maximize the system throughput, we formulate a sum rate optimization problem. The formulated problem is non-convex and complex due to the binary selection variables and the high coupling of all optimization variables. To tackle this problem, we propose an effective alternating optimization method. Specifically, fractional programming is employed to optimize the active and passive beamforming, and enumeration algorithm is used for the assisted IRS selection.

The remainder of this paper is organized as follows. System model of multi-IRS assisted mmWave system and the corresponding optimization problem are described in Section II. Section III presents the method to solve the formulated optimization problem. Simulation results are provided in Section IV to verify the performance of the proposed method. Finally, Section V concludes the paper.

\section{System Model and Problem Formulation}
\subsection{System Model}
\begin{figure}[t]
	\begin{center}
		\scalebox{0.5}[0.5]{\includegraphics{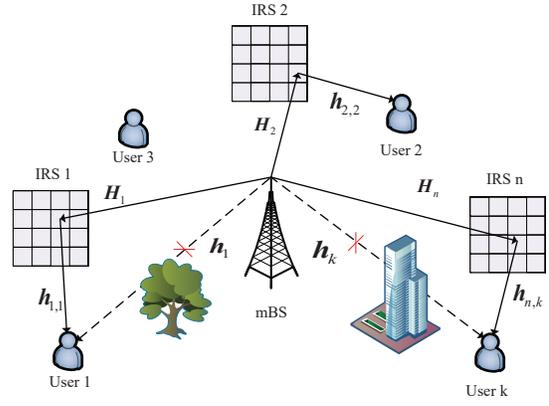}}
		\caption{A multi-IRS aided mmWave MISO system.}
		\label{fig:1}
	\end{center}
\end{figure}

A multi-IRS aided multi-user downlink mmWave multiple-input single-output (MISO) system is considered, as shown in Fig.~\ref{fig:1}, where $N$ IRSs are deployed around the mBS to assist in communication from mBS to $K$ single-antenna users. The mBS is equipped with an uniform linear array (ULA), consisting of $M$ antennas, while each IRS is modeled as an uniform planar array (UPA) with $L = {L_v} \!\times\! {L_h}$ reflecting elements, where ${L_v}$ and ${L_h}$ denote the number of reflecting elements in the vertical and horizontal directions, respectively. The index sets of IRSs, users and IRS elements are defined as ${\cal N} \buildrel \Delta \over = \left\{ {1,2,...,N} \right\}$, ${\cal K} \buildrel \Delta \over = \left\{ {1,2,...,K} \right\}$ and ${\cal L} \buildrel \Delta \over = \left\{ {1,2,...,L} \right\}$, respectively. Let $\boldsymbol{\Theta}_{n}=\operatorname{diag}\left(\theta_{n, 1}, \theta_{n, 2},..., \theta_{n, L}\right) \in \mathbb{C}^{L \times L}$ denote the reflection coefficient matrix of IRS $n$, where ${\theta _{n,l}} = \sqrt {{\eta _{n,l}}} {e^{j{\varphi _{n,l}}}}$, $\sqrt {{\eta _{n,l}}}  \in \left[ {0,1} \right]$ and ${\varphi _{n,l}} \in \left[ {0,2\pi } \right]$ denote the amplitude and phase shift associated with $l$-th element of IRS $n$, respectively. Since IRS is designed to maximize the reflected signal gain, we assume that $\sqrt {{\eta _{n,l}}}  = 1,n \in {\cal N},\forall l \in {\cal L}$, as in \cite{IRS-active-passive}, \cite{Sum-rate-1}, \cite{Association-IRS-mmWave}. Due to the high path loss of mmWave, we ignore the signals that reflected by IRSs twice and more \cite{IRS-Cell-free-network}. For simplicity, we assume that the channel state information (CSI) is fully known at mBS \cite{Sum-rate-1}, \cite{Association-IRS-mmWave}. It should be emphasized that the availability of perfect CSI in the IRS aided systems is a challenging task. But it is out of the scope of this paper. And the uplink is our future work.	

Let ${\boldsymbol{h}_k^H} \in {\mathbb{C}^{1 \times M}}$, $k \in {\cal K}$ denote the mmWave channel from mBS to user $k$, ${\boldsymbol{H}_n} \in {\mathbb{C}^{L \times M}}$, $n \in {\cal N}$ denote the mmWave channel from mBS to IRS $n$, and ${\boldsymbol{h}_{n,k}^H} \in {\mathbb{C}^{1 \times L}}$, $n \in {\cal N}$, $k \in {\cal K}$ denote the channel from IRS $n$ to user $k$, respectively. According to the Saleh-Valenzuela channel model \cite{S-V-model}, \cite{mmWave-multi-IRS}, ${\boldsymbol{h}_k}$ is expressed as
\begin{equation}
\boldsymbol{h}_{k}^H=\sqrt{\frac{M}{N_{B-U}}} \sum_{l=1}^{N_{B-U}} \rho_{l} \boldsymbol{a}_{t}\left(\phi_{l}^{\mathrm{AOD}}\right),
\end{equation}
where $N_{B-U}$ is the number of paths between the mBS and user $k$, $\rho_{l}$ is the complex gain of the $l$-th path, $\phi_{l}^{\mathrm{AOD}}$ is the associated angle of departure, and $\boldsymbol{a}_{t}$ denotes the normalized array response vector at transmitter.
\begin{small}
	\begin{equation}
	\boldsymbol{a}_t\left( {\phi _l^{\mathrm{AOD}}} \right) = \frac{1}{{\sqrt M }}\left[ {1,{e^{j\pi\sin \phi _l^{\mathrm{AOD}}}},...,{e^{j\pi \left( {M - 1} \right)\sin \phi _l^{\mathrm{AOD}}}}} \right].
	\end{equation}
\end{small}
Meanwhile, the channel between mBS and IRS $n$ is given as
\begin{equation}
\boldsymbol{H}_{n}=\sqrt{\frac{M L}{N_{B-I}}}\left(\sum_{l=1}^{N_{B-I}} \zeta_{l} \boldsymbol{a}_{r}\left(\vartheta_{al}^{\mathrm{AOA}}, \vartheta_{el}^{\mathrm{AOA}}\right) \boldsymbol{a}_{t}^{H}\left(\phi_{l}^{\mathrm{AOD}}\right)\right),
\end{equation}
where $N_{B-I}$ is the number of paths between the mBS and IRS $n$, $l = 1$ denotes the line-of-sight (LOS) path, $l > 1$ denotes the non-line-of-sight (NLOS) path, $\zeta_{l}$ is the complex gain associated with the $l$-th path, $\vartheta_{a l}^{\mathrm{AOA}}$($\vartheta_{e l}^{\mathrm{AOA}}$) represents the azimuth (elevation) angle of arrival of the $l$-th path, $\boldsymbol{a}_{r}$($\boldsymbol{a}_{t}$) is the normalized array response vector at the receiver (transmitter) as
\begin{equation}
\begin{array}{l}
{{\bf{a}}_r}\left( {\vartheta _{al}^{{\rm{AOA}}},\vartheta _{el}^{{\rm{AOA}}}} \right) = \frac{1}{{\sqrt L }}\left[ {1,{e^{j\pi \left( {\sin \vartheta _{al}^{{\rm{AOA}}}\sin \vartheta _{el}^{{\rm{AOA}}} + \cos \vartheta _{el}^{{\rm{AOA}}}} \right)}},} \right.\\
\ldots ,{\left. {{e^{j\pi \left( {\left( {{L_v} - 1} \right)\sin \vartheta _{al}^{{\rm{AOA}}}\sin \vartheta _{el}^{{\rm{AOA}}} + \left( {{L_h} - 1} \right)\cos \vartheta _{el}^{{\rm{AOA}}}} \right)}}} \right]^T}.
\end{array}
\end{equation}

Moreover, the channel from IRS $n$ to user $k$ can be expressed as
\begin{equation}
\boldsymbol{h}_{n,k}^H = \sqrt {\frac{L}{{{N_{I - k}}}}} \left( {\sum\limits_{l = 1}^{{N_{I - k}}} {{\varrho _l}} {{\bf{a}}_t}\left( {\vartheta _{al}^{\mathrm{AOD}},\vartheta _{el}^{\mathrm{AOD}}} \right)} \right),
\end{equation}
where $N_{I-k}$ is the number of path from IRS $n$ to user $k$, $\varrho _l$ is the complex gain associated with the $l$-th path, and $\vartheta _{al}^{\mathrm{AOD}}$($\vartheta _{el}^{\mathrm{AOD}}$) is the azimuth (elevation) angle of departure associated with the $l$-th path.

Let $\boldsymbol{x} = \sum\limits_{k = 1}^K {{\boldsymbol{p}_k}{s_k}}$ denote the transmitted signal at mBS, where $s_k$ is the transmit symbol for user $k$ and $\boldsymbol{p}_k \in {\mathbb{C}^{M \times 1}}$ is the corresponding beamforming vector. In general, the received signal of user $k$ can be expressed as
\begin{equation}
y_{k}=\left(\boldsymbol{h}_{k}^H+\sum_{n \in \mathcal{N}} \boldsymbol{h}_{n, k}^H \boldsymbol{\Theta}_{n}^H \boldsymbol{H}_{n}\right) \sum_{i \in \mathcal{K}} \boldsymbol{p}_{i} s_{i}+n_{k}, \label{6}
\end{equation}
where $n_{k}$ is the complex additive white Gaussian noise, ${n_k} \sim {\cal C}{\cal N}\left( {0,{\sigma ^2}} \right)$.

In this paper, we assume that the direct links from the mBS to users are severely obstacled by buildings or other things. Meanwhile, to simplify the design of the reflection matrix at IRS, we assume that each IRS only serves maximal one user at a moment, and it serves multiple users in a time-division manner. For the IRS selection, we define the binary variables ${a_{n,k}}\!\in\! \left\{{0,1} \right\},n \in {\cal N},k \in {\cal K}$, where ${a_{n,k}} = 1$ indicates that user $k$ is served by IRS $n$, otherwise we have ${a_{n,k}} = 0$. As such, for IRS $n$, we have $\sum\limits_{k \in {\cal K}} {{a_{n,k}} = 1} ,\forall n \in {\cal N}$, and the IRS selection matrix $\boldsymbol{A} \in \mathbb{C}^{N \times K}$ can be written as
\begin{equation}
\boldsymbol{A} = \left[ {\begin{array}{*{20}{c}}
	{{a_{1,1}}}& \cdots &{{a_{1,K}}}\\
	\vdots & \ddots & \vdots \\
	{{a_{N,1}}}& \cdots &{{a_{N,K}}}
	\end{array}} \right].
\end{equation}

\subsection{Problem Formulation}
The signal-to-interference-plus-noise ratio (SINR) of user $k$ is given by
\begin{equation}
\operatorname{SINR}_{k}=\frac{\left|\sum_{n \in \mathcal{N}} a_{n, k}\left(\boldsymbol{h}_{n, k}^H \boldsymbol{\Theta}_{n}^H \boldsymbol{H}_{n}\right) \boldsymbol{p}_{k}\right|^{2}}{\sum_{\substack{i \in \mathcal{K} \\ i \neq k}}\left|\left(\sum_{n \in \mathcal{N}} a_{n, k}\left(\boldsymbol{h}_{n, k}^H \boldsymbol{\Theta}_{n}^H \boldsymbol{H}_{n}\right)\right) \boldsymbol{p}_{i}\right|^{2}+\sigma^{2}}. \label{SINR_k}
\end{equation}
Thereby, the achievable rate (bps/Hz) for user $k$ can be expressed as
\begin{equation}
R_{k}=\log _{2}\left(1+\operatorname{SINR}_{k}\right).
\end{equation}

In this paper, we aim to maxmize the sum rate of all users by jointly optimizing active and passive beamforming, subject to the constraint of SINR at each user and the constraint of IRS selection. Then the problem can be formulated as

\begin{align}
(\mathrm{P1}) \max _{\boldsymbol{P}, \boldsymbol{\Theta}_{n}, \boldsymbol{A}} &~ f_{1}\left(\boldsymbol{P}, \boldsymbol{\Theta}_{n}, \boldsymbol{A}\right)=\sum_{k \in \mathcal{K}} R_{k}, \label{equ:P1} \\
{\rm s.t.}~	
& \sum_{k \in \mathcal{K}}\left\|\boldsymbol{p}_{k}\right\|^{2} \leq P_{\max }, \tag{\ref{equ:P1}a}\\
&\left|\theta_{n, l}\right|^{2}=1, \forall n \in {\cal N}, \forall l \in {\cal L}, \tag{\ref{equ:P1}b} \\
& \operatorname{SINR}_{k} \geq \operatorname{SINR}_{\min }, \forall k \in {\cal K}, \tag{\ref{equ:P1}c} \\
& \sum_{k \in \mathcal{K}} a_{n, k}=1, \forall n \in {\cal N}, \tag{\ref{equ:P1}d}\\
& a_{n, k} \in\{0,1\}, \forall n \in {\cal N}, \forall k \in {\cal K}, \tag{\ref{equ:P1}e}
\end{align}
where $\boldsymbol{P} = {\left[ {{\boldsymbol{p}_1}^T,{\boldsymbol{p}_2}^T, \cdots ,{\boldsymbol{p}_K}^T} \right]^T}$, constraint (\ref{equ:P1}a) limits the total transmit power of the mBS, constraint (\ref{equ:P1}b) is the reflection coefficient constraint for passive beamforming, and constraint (\ref{equ:P1}c) denotes the minimum SINR for reliable communincation.  Problem $(\mathrm{P1})$ is a complex non-convex problem due to the non-convex objective function (\ref{equ:P1}) and the non-convex constraints (\ref{equ:P1}b)-(\ref{equ:P1}e). In next section, we will explore how to effectively solve this problem.

\section{Problem Solution}
In this section, we propose an alternative optimization method to solve peoblem $(\mathrm{P1})$. Firstly, an equivalent problem of $(\mathrm{P1})$ is established by employing the Lagrangian dual transform proposed in \cite{Fractional-Programming}. Then, we decouple it into three sub-optimization problems, where the active and passive beamfroming are optimized by fractional programming and the IRS selection is optimized via enumeration algorithm.

\subsection{Equivalent Transformation of (P1)}
We utilize the Lagrangian dual transform \cite{Fractional-Programming} to obtain an equivalent problem of $(\mathrm{P1})$. By introducing an auxiliary variable $\boldsymbol{\alpha}=\left[\alpha_{1}, \alpha_{2}, \cdots, \alpha_{K}\right]^{T}$, the objective function (\ref{equ:P1}) can be equivalently expressed as
\begin{equation}
\begin{aligned}
&f_{1 a}\left(\boldsymbol{P}, \boldsymbol{\Theta}_{n}, \boldsymbol{A}, \boldsymbol{\alpha}\right) \\
&=\sum_{k \in \mathcal{K}} \log _{2}\left(1+\alpha_{k}\right)-\sum_{k \in \mathcal{K}} \frac{\alpha_{k}}{\ln 2}+\sum_{k \in \mathcal{K}} \frac{\left(1+\alpha_{k}\right) \mathrm{SINR}_{k}}{\left(1+\mathrm{SINR}_{k}\right) \ln 2}.
\end{aligned}
\end{equation}
Hence, problem $(\mathrm{P1})$ can be reformulated as
\begin{align}
(\mathrm{P1.1}) \max _{\boldsymbol{P}, \boldsymbol{\Theta}_{n}, \boldsymbol{A}, \boldsymbol{\alpha}} &~ f_{1a}\left(\boldsymbol{P}, \boldsymbol{\Theta}_{n}, \boldsymbol{A}, \boldsymbol{\alpha}\right), \label{equ:P1.1} \\
{\rm s.t.}~
&(10 \mathrm{a})-(10 \mathrm{e}). \tag{\ref{equ:P1.1}a}
\end{align}
In problem $(\mathrm{P} 1.1)$, when $\boldsymbol{P}$, $\boldsymbol{\Theta}_{n}$ and $\boldsymbol{A}$ are fixed, the optimal $\alpha_k$, denoted by $\alpha _k^{opt}$, can be obtained by setting ${{\partial {f_{1a}}} \mathord{\left/
		{\vphantom {{\partial {f_2}} {\partial {\alpha_k}}}} \right.
		\kern-\nulldelimiterspace} {\partial {\alpha_k}}}$ to zero. We have $\alpha _k^{opt} = \mathrm{SINR}_{k}$.

Note that given $\alpha _k$, only the last term of $f_{1a}$ is connected with $\boldsymbol{P}$, $\boldsymbol{\Theta}_{n}$ and $\boldsymbol{A}$. Consequently, the original problem can be further simplified as
\begin{align}
(\mathrm{P1.2}) \max _{\boldsymbol{P}, \boldsymbol{\Theta}_{n}, \boldsymbol{A}} &~ f_{1b}\left(\boldsymbol{P}, \boldsymbol{\Theta}_{n}, \boldsymbol{A}\right)=\sum_{k \in \mathcal{K}} \frac{\left(1+\alpha_{k}\right) \mathrm{SINR}_{k}}{\left(1+\mathrm{SINR}_{k}\right)}, \label{equ:P1.2}\\
{\rm s.t.}~
& (10 \mathrm{a})-(10 \mathrm{e}). \tag{\ref{equ:P1.2}a}
\end{align}

The above problem $(\mathrm{P1.2})$ can be further solved as shown in subsection III-B, III-C and III-D.

\subsection{Active Beamforming Optimization at mBS}
To facilitate the description, we define $\hat{\boldsymbol{h}}_{n, k}^H=\boldsymbol{h}_{n, k}^H \boldsymbol{\Theta}_{n}^H \boldsymbol{H}_{n}$. Substituting $\hat{\boldsymbol{h}}_{n, k}^H$ and (\ref{SINR_k}) into (\ref{equ:P1.2}), we rewrite $f_{1b}$ with given $\boldsymbol{\alpha}$, $\boldsymbol{\Theta}_{n}$ and $\boldsymbol{A}$ as
\begin{equation}
\begin{aligned} \label{f2}
f_{2}(\boldsymbol{P}) &=\sum_{k \in \mathcal{K}} \frac{\left(1+\alpha_{k}\right) \operatorname{SINR}_{k}}{\left(1+\operatorname{SINR}_{k}\right)} \\
&=\sum_{k \in \mathcal{K}} \frac{\left(1+\alpha_{k}\right)\left|\left(\sum_{n \in \mathcal{N}} a_{n, k} \hat{\boldsymbol{h}}_{n, k}^H\right) \boldsymbol{p}_{k}\right|^{2}}{\sum_{i \in \mathcal{K}}\left|\left(\sum_{n \in \mathcal{N}} a_{n, k} \hat{\boldsymbol{h}}_{n, k}^H\right) \boldsymbol{p}_{i}\right|^{2}+\sigma^{2}}.
\end{aligned}
\end{equation}
Thereby, the problem of optimizing $\boldsymbol{P}$ can be expressed as
\begin{align}
(\mathrm{P2}) \max _{\boldsymbol{P}} &~f_{2}(\boldsymbol{P}), \label{equ:P2}\\
{\rm s.t.}~
& (10 \mathrm{a}),(10 \mathrm{c}). \tag{\ref{equ:P2}a}
\end{align}
We see that $f_{2}$ is in a sum-of-ratio form, which can be tackled by applying quadratic transform \cite{Fractional-Programming} as follows
\begin{align}
(\mathrm{P2.1}) \max _{\boldsymbol{P}, \boldsymbol{\varepsilon}} &~ f_{2a}(\boldsymbol{P},\boldsymbol{\varepsilon}), \label{equ:P2.1}\\
{\rm s.t.}~
& (10 \mathrm{a}),(10 \mathrm{c}), \tag{\ref{equ:P2.1}a}
\end{align}
where $\boldsymbol{\varepsilon}=\left[\mathcal{\varepsilon}_{1}, \mathcal{\varepsilon}_{2}, \cdots, \mathcal{\varepsilon}_{K}\right]^{T}$ is an auxiliary variable and the objective function $f_{2a}(\boldsymbol{P},\boldsymbol{\varepsilon})$ is denoted by
\begin{equation}
\begin{aligned}
f_{2a}(\boldsymbol{P}, \boldsymbol{\varepsilon})=& \sum_{k \in \mathcal{K}} 2 \sqrt{\left(1+\alpha_{k}\right)} \mathfrak{R}\left\{\varepsilon_{k}^{*} \boldsymbol{H}_{k}^H \boldsymbol{p}_{k}\right\} \\
&-\sum_{k \in \mathcal{K}}\left|\varepsilon_{k}\right|^{2}\left(\sum_{i \in \mathcal{K}}\left|\boldsymbol{H}_{k}^H \boldsymbol{p}_{i}\right|^{2}+\sigma^{2}\right),
\end{aligned}
\end{equation}
and $\boldsymbol{H}_{k}^H=\sum_{n \in \mathcal{N}} a_{n, k} \hat{\boldsymbol{h}}_{n, k}^H$.

Given $\boldsymbol{P}$, the optimal $\mathcal{\varepsilon}_{k}$ is obtained by setting ${{\partial f_{2a}} \mathord{\left/
		{\vphantom {{\partial {f_5}} {\partial {\varepsilon_k}}}} \right.
		\kern-\nulldelimiterspace} {\partial {\varepsilon_k}}}$ to zero as
\begin{equation}
\varepsilon_{k}^{o p t}=\frac{\sqrt{\left(1+\alpha_{k}\right)} \boldsymbol{H}_{k}^H \boldsymbol{p}_{k}}{\left(\sum_{i \in \mathcal{K}}\left|\boldsymbol{H}_{k}^H \boldsymbol{p}_{i}\right|^{2}+\sigma^{2}\right)}.
\end{equation}
The following problem is optimizing $\boldsymbol{P}$ with given $\boldsymbol{\varepsilon}$. Let $\mathbf{v}_{k}=\sqrt{\left(1+\alpha_{k}\right)} \varepsilon_{k} \boldsymbol{H}_{k}$, $\mathbf{z}=\sum_{k \in \mathcal{K}}\left|\varepsilon_{k}\right|^{2} \boldsymbol{H}_{k} \boldsymbol{H}_{k}^{H}$ and $g = \sum\limits_{k \in {\cal K}} {{{\left| {{\varepsilon _k}} \right|}^2}{\sigma ^2}} $. The $f_{2a}$ can be simplified as
\begin{equation}
f_{2b}(\boldsymbol{P})=-\boldsymbol{P}^{H} \mathbf{Z} \boldsymbol{P}+2 \mathfrak{R}\left\{\mathbf{V}^{H} \boldsymbol{P}\right\}-g,
\end{equation}
where $\mathbf{Z} = \operatorname{diag}\left( {{\mathbf{z}_1}, \cdots ,{\mathbf{z}_K}} \right)$, $\mathbf{z}_{k}=\mathbf{z}, \forall k \in \mathcal{K}$ and $\mathbf{V} = {\left[ {{\mathbf{v}_1^T}, \cdots ,{\mathbf{v}_K^T}} \right]^T}$.

Thereby, problem $(\mathrm{P2.1})$ can be further simplified as
\begin{align}
(\mathrm{P2.2}) \max _{\boldsymbol{P}} &~ f_{2b}(\boldsymbol{P})=-\boldsymbol{P}^{H} \mathbf{Z} \boldsymbol{P}+2 \mathfrak{R}\left\{\mathbf{V}^{H} \boldsymbol{P}\right\}-g, \label{equ:P2.2}\\
{\rm s.t.}~
& (10 \mathrm{a}),(10 \mathrm{c}). \tag{\ref{equ:P2.2}a}
\end{align}
Although the objective function $f_{2b}$ and constraint $(10 \mathrm{a})$ are convex, the problem $(\mathrm{P2.2})$ is non-convex due to the non-convex constraint $(10 \mathrm{c})$.

According to \cite{SINR-SOCP}, by using second-order-cone programming (SOCP), we rewrite constraint $(10 \mathrm{c})$ as
\begin{equation} \label{22}
\sqrt {1 + \frac{1}{{\operatorname{SINR}{_{\min }}}}} {\boldsymbol{H}_k^H}{\boldsymbol{p}_k} \ge \left\| {\begin{array}{*{20}{c}}
	{{\boldsymbol{H}_k^H}\boldsymbol{\bar P}}\\
	\sigma
	\end{array}} \right\|,
\end{equation}
\begin{equation} \label{23}
{\boldsymbol{H}_k^H}{\boldsymbol{p}_k} \ge 0,\forall k \in {\cal K},
\end{equation}
where $\boldsymbol{\bar P} = {\left[ {{\boldsymbol{p}_1},{\boldsymbol{p}_2}, \cdots ,{\boldsymbol{p}_K}} \right]}$. Therefore, the problem of active beamforming optimization can be eventually solved as following form
\begin{align}
(\mathrm{P2.3}) \max _{\boldsymbol{P}} &~ f_{2b}(\boldsymbol{P})=-\boldsymbol{P}^{H} \mathbf{Z} \boldsymbol{P}+2 \mathfrak{R}\left\{\mathbf{V}^{H} \boldsymbol{P}\right\}-g, \label{equ:P2.3}\\
{\rm s.t.}~
& (10 \mathrm{a}),(21) \text { and } (22). \tag{\ref{equ:P2.3}a}
\end{align}
Problem $(\mathrm{P2.3})$ is convex, which can be solved by standard convex optimization solvers such as CVX. Hence, the optimal $\boldsymbol{P}$ is obtained by solving $(\mathrm{P2.3})$.

\subsection{Passive Beamforming Optimization at IRS}
First of all, we rewrite $\hat{\boldsymbol{h}}_{n, k}^H=\boldsymbol{h}_{n, k}^H \boldsymbol{\Theta}_{n}^H \boldsymbol{H}_{n}$ as
\begin{equation}
\hat{\boldsymbol{h}}_{n, k}^H \boldsymbol{p}_{i}=\left(\boldsymbol{h}_{n, k}^H \boldsymbol{\Theta}_{n}^H \boldsymbol{H}_{n}\right) \boldsymbol{p}_{i}=\boldsymbol{\theta}_{n}^{H} \operatorname{diag}\left(\boldsymbol{h}_{n, k}^H\right) \boldsymbol{H}_{n} \boldsymbol{p}_{i},
\end{equation}
where $\boldsymbol{\theta}_{n}=\left[\theta_{n, 1}, \cdots, \theta_{n, L}\right]^{T}$.

Let $\boldsymbol{b}_{n, k, i}=\operatorname{diag}\left(\boldsymbol{h}_{n, k}^H\right) \boldsymbol{H}_{n} \boldsymbol{p}_{i}$, then substituting $\boldsymbol{\theta}_{n}$ and $\boldsymbol{b}_{n, k, i}$ into (\ref{f2}), $f_{2}$ is rewritten as
\begin{equation}
f_{3}\left(\boldsymbol{\theta}_{n}\right)=\sum_{k \in \mathcal{K}} \frac{\left(1+\alpha_{k}\right)\left|\sum_{n \in \mathcal{N}} a_{n, k} \boldsymbol{\theta}_{n}^{H} \boldsymbol{b}_{n, k, k}\right|^{2}}{\sum_{i \in \mathcal{K}}\left|\sum_{n \in \mathcal{N}} a_{n, k} \boldsymbol{\theta}_{n}^{H} \boldsymbol{b}_{n, k, i}\right|^{2}+\sigma^{2}}.
\end{equation}
We define $\boldsymbol{\Theta}=\left[\boldsymbol{\theta}_{1}^T, \cdots, \boldsymbol{\theta}_{N}^T\right]^{T}$ and $\boldsymbol{B}_{k,i}=\left[a_{1, k} \boldsymbol{b}_{1, k, i}^T, \cdots, a_{N, k} \boldsymbol{b}_{N, k, i}^T\right]^{T}$. Thereby, $f_{3}$ is equivalently transformed to a new function of $\boldsymbol{\Theta}$
\begin{equation}
f_{3a}(\boldsymbol{\Theta})=\sum_{k \in \mathcal{K}} \frac{\left(1+\alpha_{k}\right)\left|\boldsymbol{\Theta}^{H} \boldsymbol{B}_{k,k}\right|^{2}}{\sum_{i \in \mathcal{K}}\left|\boldsymbol{\Theta}^{H} \boldsymbol{B}_{k,i}\right|^{2}+\sigma^{2}}.
\end{equation}
Consequently, given $\boldsymbol{\alpha}$, $\boldsymbol{P}$ and $\boldsymbol{A}$, the passive beamforming optimization problem is expressed as
\begin{align}
(\mathrm{P3}) \max _{\boldsymbol{\Theta}} &~f_{3a}(\boldsymbol{\Theta}), \label{equ:P3}\\
{\rm s.t.}~
& (10 \mathrm{b}),(10 \mathrm{c}). \tag{\ref{equ:P3}a}
\end{align}

Similarly, introducing auxiliary variable $\boldsymbol{\beta}=\left[\beta_{1}, \cdots, \beta_{K}\right]^{T}$, $f_{3a}$ is transformed as
\begin{equation}
\begin{aligned}
f_{3b}(\boldsymbol{\Theta}, \boldsymbol{\beta})=& \sum_{k \in \mathcal{K}} 2 \sqrt{\left(1+\alpha_{k}\right)} \mathfrak{R}\left\{\beta_{k}^{*} \boldsymbol{\Theta}^{H} \boldsymbol{B}_{k,k}\right\} \\
&-\sum_{k \in \mathcal{K}}\left|\beta_{k}\right|^{2}\left(\sum_{i \in \mathcal{K}}\left|\boldsymbol{\Theta}^{H} \boldsymbol{B}_{k,i}\right|^{2}+\sigma^{2}\right).
\end{aligned}
\end{equation}
Let ${{\partial f_{3b}} \mathord{\left/
		{\vphantom {{\partial f_{3b}} {\partial {\beta _k}}}} \right.
		\kern-\nulldelimiterspace} {\partial {\beta _k}}}$ to zero, the optimal $\beta _k$ is obtained as
\begin{equation}
\beta_{k}^{o p t}=\frac{\sqrt{\left(1+\alpha_{k}\right)}\left(\boldsymbol{\Theta}^{H} \boldsymbol{B}_{k,k}\right)}{\sum_{i \in \mathcal{K}}\left|\boldsymbol{\Theta}^{H} \boldsymbol{B}_{k,i}\right|^{2}+\sigma^{2}}.
\end{equation}

Given $\boldsymbol{\beta}$, the optimization of $\boldsymbol{\Theta}$ can be represented as follows
\begin{align}
(\mathrm{P3.1}) \max _{\boldsymbol{\Theta}} &~f_{3c}(\boldsymbol{\Theta})=-\boldsymbol{\Theta}^{H} \boldsymbol{U} \boldsymbol{\Theta}+2 \mathfrak{R}\left\{\boldsymbol{\Theta}^{H} \boldsymbol{D}\right\}-c, \label{equ:P3.1}\\
{\rm s.t.}~
& (10 \mathrm{b}),(10 \mathrm{c}), \tag{\ref{equ:P3.1}a}
\end{align}
where $\boldsymbol{U} = \sum\limits_{k \in {\cal K}} {{{\left| {{\beta _k}} \right|}^2}\left( {\sum\limits_{i \in {\cal K}} {{\boldsymbol{B}_{k,i}}\boldsymbol{B}_{k,i}^H} } \right)} $, $c = \sum\limits_{k \in {\cal K}} {{{\left| {{\beta _k}} \right|}^2}{\sigma ^2}} $ and $\boldsymbol{D}=\sum_{k \in \mathcal{K}} \sqrt{\left(1+\alpha_{k}\right)} \beta_{k}^{*} \boldsymbol{B}_{k,k}$.

Since the constraints $(10 \mathrm{b})$ and $(10 \mathrm{c})$ are non-convex, the problem $(\mathrm{P3.1})$ is non-convex. For constraint $(10 \mathrm{b})$, it is an unit-modulus constraint. We relax constraint $(10 \mathrm{b})$, i.e. $\left|\theta_{n, l}\right|^{2} \le 1, \forall n \in {\cal N}, \forall l \in {\cal L}$. For constraint $(10 \mathrm{c})$, similar to (\ref{22}) and (\ref{23}), we handle it with SOCP method. Then the new constraint is denoted by
\begin{equation}
\sqrt {\left( {1 + \frac{1}{{\operatorname{SINR}{_{\min }}}}} \right)} {\boldsymbol{\Theta} ^H}{\boldsymbol{B}_{k,k}} \ge \left\| {\begin{array}{*{20}{c}}
	{{\boldsymbol{\Theta} ^H}\boldsymbol{B}_{k}}\\
	\sigma
	\end{array}} \right\|,
\end{equation}
\begin{equation}
{\boldsymbol{\Theta} ^H}{\boldsymbol{B}_{k,k}} \ge 0,\forall k \in {\cal K},
\end{equation}
where $\boldsymbol{B}_{k} = {\left[ {{\boldsymbol{B}_{k,1}},{\boldsymbol{B}_{k,2}}, \cdots ,{\boldsymbol{B}_{k,K}}} \right]}$.

Finally, the passive beamforming optimization is reformulated as
\begin{align}
(\mathrm{P3.2}) \max _{\boldsymbol{\Theta}} &~f_{3c}(\boldsymbol{\Theta})=-\boldsymbol{\Theta}^{H} \boldsymbol{U} \boldsymbol{\Theta}+2 \mathfrak{R}\left\{\boldsymbol{\Theta}^{H} \boldsymbol{D}\right\}-c, \label{equ:P3.2}\\
{\rm s.t.}~
& \left|\theta_{n, l}\right|^{2} \le 1, \forall n \in {\cal N}, \forall l \in {\cal L}, \tag{\ref{equ:P3.2}a}\\
& (31), (32), \tag{\ref{equ:P3.2}b}
\end{align}
which is convex and can be solve by CVX.

\subsection{IRS selection Optimization}
Given $\boldsymbol{\alpha}$, $\boldsymbol{P}$ and $\boldsymbol{\Theta}$, the IRS selection optimization problem can be expressed as
\begin{align}
(\mathrm{P4}) \max _{\boldsymbol{A}} &~ f_{4}\left(\boldsymbol{A}\right)=\sum_{k \in \mathcal{K}} R_{k}, \label{equ:P4} \\
{\rm s.t.}~	
& (10 \mathrm{d}), (10 \mathrm{e}). \tag{\ref{equ:P4}a}
\end{align}
Due to $a_{n, k} \in\{0,1\}, \forall n \in {\cal N}, \forall k \in {\cal K}$, and the dimension of matrix $\boldsymbol{A}$ will not be very large, we can obtain the optimal $\boldsymbol{A}$ by enumerating all possible combinations between IRSs and users. In the worst case, the complexity of enumeration is $\mathcal{O}\left(K^{N}\right)$.

Finally, we update $\boldsymbol{\alpha}$, $\boldsymbol{P}$, $\boldsymbol{\Theta}$ and $\boldsymbol{A}$ in an alternate manner, until problem $(\mathrm{P1})$ reaches a stable optimal solution.

\section{Simulation and Discussions}
To verify the performance of the proposed method, we consider a simulation scenario that the mBS is located at the origin, $N=6$ IRSs are located at $(60{\rm{m}},40{\rm{m}})$, $(60{\rm{m}},-40{\rm{m}})$, $(100{\rm{m}},40{\rm{m}})$, $(100{\rm{m}},-40{\rm{m}})$, $(140{\rm{m}},40{\rm{m}})$ and $(140{\rm{m}},-40{\rm{m}})$, respectively. The users are randomly distributed in a circle at $(100{\rm{m}},0{\rm{m}})$ with a radius of $10{\rm{m}}$. The number of antenna at mBS $M=40$. According to \cite{Association-IRS-mmWave} and \cite{S-V-model}, the channel gain is generated as $\rho_{l} \sim \mathcal{C N}\left(0,10^{-0.1 \kappa}\right)$ and $\kappa=a+10 b \log _{10}(d)+\xi$, where $\xi \sim \mathcal{N}\left(0, \sigma_{\xi}^{2}\right)$. For LOS path, the values of $a$, $b$ and $\sigma_{\xi}$  are respectively setting as $a=61.4$, $b=2$ and $\sigma_{\xi}=5.8$ dB. For NLOS path, the values of $a$, $b$ and $\sigma_{\xi}$  are setting as $a=72$, $b=2.92$ and $\sigma_{\xi}=8.7$ dB, respectively. The generation of channel complex gains $\zeta_{l}$ and $\varrho _l$  are similar to $\rho_{l}$. And other parameters are set as follows: $N_{B-I} = 5$, $N_{I-k} = 1$, ${P_{\max }} = 30$ dBm, ${\rm{SIN}}{{\rm{R}}_{\min }} =  - 20$ dB and ${\sigma ^2} =  - 90$ dBm.

\begin{figure}[t]
	\begin{center}
		\scalebox{0.5}[0.5]{\includegraphics{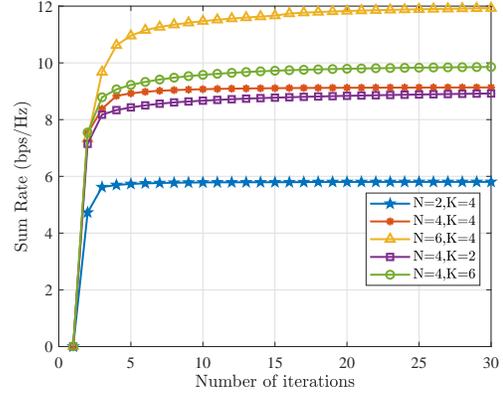}}
		\caption{Convergence performance of the proposed method.}
		\label{fig:2}
	\end{center}
\end{figure}
\begin{figure}[t]
	\begin{center}
		\scalebox{0.5}[0.5]{\includegraphics{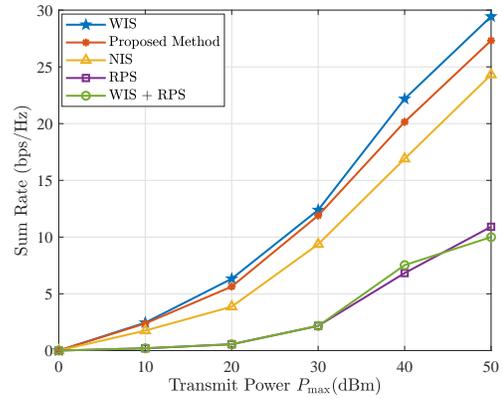}}
		\caption{The sum rate versus ${P_{\max }}$ with $N=4$, $K=4$.}
		\label{fig:3}
	\end{center}
\end{figure}

In order to evaluate the proposed scheme, we compare it with the following three methods: 1) WIS: without consider the IRS selection; 2) RPS: a random passive beamforming is used at IRS; 3) NIS: select the nearest IRS for service.

To show the convergence of the proposed method, we plot the sum rate against the number of iterations under different number of IRSs and users in Fig.~\ref{fig:2}. We can see that with the increase of the number of IRSs $N$ and the number of users $K$, the convergence speed becomes slower but the sum rate obtains significant improvement. For example, when $N=K=4$, the optimal solution is obtained within 10 iterations and the sum rate of convergence is about 9 bps/Hz, while when $N=6$ and $K=4$, the optimal solution needs about 25 iterations to converge and the corresponding rate can up to about 12 bps/Hz. Besides, we notice that when $N$ is fixed, increasing $K$ has little effect on the sum rate. But when $K$ is fixed, increasing $N$ will greatly improve the rate performance. The reason is that we introduced the IRS selection in the proposed method, so that the number of IRSs largely determines the number of users that the system can serve simultaneously.

\begin{figure}[t]
	\begin{center}
		\scalebox{0.5}[0.5]{\includegraphics{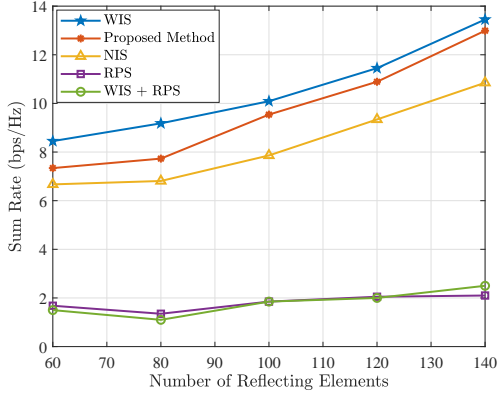}}
		\caption{The sum rate versus $L$ with $N=4$, $K=4$.}
		\label{fig:4}
	\end{center}
\end{figure}

Fig.~\ref{fig:3} shows the impact of the maximum transmit power on the sum rate performance. This figure shows that the sum rate increases with the transmit power and the proposed method outperforms all other methods except the ``WIS''.  For example, when $P_{\max}= 50$ dBm, the sum rate of the proposed method can reach about 27 bps/Hz, which is about 3 bps/Hz higher than the ``NIS''. Comparing the proposed method with the ``WIS'', we observe that when $P_{\max} \leq 30$ dBm, the performance gap between them is very small but gradually becomes larger when $P_{\max}$ exceeds $30$ dBm. The main reason is that, when the value of $P_{\max}$ is small, the reflected signals received by the users far away from the IRS will be very weak and contribute little to the gain of the sum rate. Although the IRS selection results in a slight performance degradation compared to the ``WIS'', the IRS selection simplifies the design of the IRS phase shift matrix while providing higher data rate for the user served.

The impact of the number of reflecting elements at each IRS is shown in Fig.~\ref{fig:4}. In general, the sum rate of the ``WIS'', the ``Proposed Method'' and the ``NIS'' increases with the number of reflecting elements $L$. And the sum rate of the other two methods barely increases when the reflecting unit of IRS is increased. The reason is that the more reflective elements, the stronger the reflected signals, and the more effective the passive beamforming will be, so that the system throughput can be improved.

\section{Conclusions}
To explore utilizing IRS to tackle the problem of seriously path loss and blockage of mmWave, a multi-IRS assisted mmWave communication system is considered and we investigate a sum rate maximization problem by jointly optimizing active and passive beamforming as well as the optimal IRS selection. To tackle the challenging problem, an alternating iterative approach is proposed, where the transmit beamforming of the mBS, the passive beamforming of the IRS and the IRS selection matrix are updated in an alternating manner. The simulation results show the feasibility and effectiveness of the proposed method.

\ifCLASSOPTIONcaptionsoff
  \newpage
\fi

\bibliographystyle{IEEEtran}
\bibliography{reference}

\begin{thebibliography}{10}
\providecommand{\url}[1]{#1}
\csname url@samestyle\endcsname
\providecommand{\newblock}{\relax}
\providecommand{\bibinfo}[2]{#2}
\providecommand{\BIBentrySTDinterwordspacing}{\spaceskip=0pt\relax}
\providecommand{\BIBentryALTinterwordstretchfactor}{4}
\providecommand{\BIBentryALTinterwordspacing}{\spaceskip=\fontdimen2\font plus
\BIBentryALTinterwordstretchfactor\fontdimen3\font minus
  \fontdimen4\font\relax}
\providecommand{\BIBforeignlanguage}[2]{{%
\expandafter\ifx\csname l@#1\endcsname\relax
\typeout{** WARNING: IEEEtran.bst: No hyphenation pattern has been}%
\typeout{** loaded for the language `#1'. Using the pattern for}%
\typeout{** the default language instead.}%
\else
\language=\csname l@#1\endcsname
\fi
#2}}
\providecommand{\BIBdecl}{\relax}
\BIBdecl

\bibitem{MIMO-OTFS}
X.~Wu, S.~Ma, and X.~Yang, ``Tensor-based low-complexity channel estimation for
  mmwave massive {MIMO-OTFS} systems,'' \emph{Journal of Communications and
  Information Networks}, vol.~5, no.~3, pp. 324--334, 2020.

\bibitem{mmWave-MIMO}
S.~A. Busari, K.~M.~S. Huq, S.~Mumtaz, L.~Dai, and J.~Rodriguez,
  ``Millimeter-wave massive {MIMO} communication for future wireless systems: A
  survey,'' \emph{IEEE Communications Surveys $\&$ Tutorials}, vol.~20, no.~2,
  pp. 836--869, 2018.

\bibitem{UDN}
M.~Kamel, W.~Hamouda, and A.~Youssef, ``Ultra-dense networks: A survey,''
  \emph{IEEE Communications Surveys $\&$ Tutorials}, vol.~18, no.~4, pp.
  2522--2545, 2016.

\bibitem{UDN-2}
Q.~Xue, Y.~Sun, J.~Wang, G.~Feng, L.~Yan, and S.~Ma, ``User-centric association
  in ultra-dense mm{W}ave networks via deep reinforcement learning,''
  \emph{IEEE Commun. Lett.}, vol.~25, no.~11, pp. 3594--3598, 2021.

\bibitem{IRS-6G}
C.~Pan, H.~Ren, K.~Wang, J.~F. Kolb, M.~Elkashlan, M.~Chen, M.~Di~Renzo,
  Y.~Hao, J.~Wang, A.~L. Swindlehurst, X.~You, and L.~Hanzo, ``Reconfigurable
  intelligent surfaces for {6G} systems: Principles, applications, and research
  directions,'' \emph{IEEE Communications Magazine}, vol.~59, no.~6, pp.
  14--20, 2021.

\bibitem{LISA}
Y.-C. Liang, R.~Long, Q.~Zhang, J.~Chen, H.~V. Cheng, and H.~Guo, ``Large
  intelligent surface/antennas ({LISA}): Making reflective radios smart,''
  \emph{Journal of Communications and Information Networks}, vol.~4, no.~2, pp.
  40--50, 2019.

\bibitem{IRS-Cell-free-network}
Z.~Zhang and L.~Dai, ``A joint precoding framework for wideband reconfigurable
  intelligent surface-aided cell-free network,'' \emph{IEEE Transactions on
  Signal Processing}, vol.~69, pp. 4085--4101, 2021.

\bibitem{IRS-active-passive}
Q.~Wu and R.~Zhang, ``Intelligent reflecting surface enhanced wireless network
  via joint active and passive beamforming,'' \emph{IEEE Transactions on
  Wireless Communications}, vol.~18, no.~11, pp. 5394--5409, 2019.

\bibitem{IRS-MIMO-up}
K.~Xu, J.~Zhang, X.~Yang, S.~Ma, and G.~Yang, ``On the sum-rate of
  {RIS}-assisted {MIMO} multiple-access channels over spatially correlated
  rician fading,'' \emph{IEEE Transactions on Communications}, vol.~69, no.~12,
  pp. 8228--8241, 2021.

\bibitem{CSIT}
J.~Zhang, J.~Liu, S.~Ma, C.-K. Wen, and S.~Jin, ``Large system achievable rate
  analysis of {RIS}-assisted {MIMO} wireless communication with statistical
  {CSIT},'' \emph{IEEE Transactions on Wireless Communications}, vol.~20,
  no.~9, pp. 5572--5585, 2021.

\bibitem{Sum-rate-1}
H.~Guo, Y.-C. Liang, J.~Chen, and E.~G. Larsson, ``Weighted sum-rate
  maximization for intelligent reflecting surface enhanced wireless networks,''
  in \emph{2019 IEEE GLOBECOM}, 2019, pp. 1--6.

\bibitem{Sum-rate-3}
S.~Gong, S.~Ma, Z.~Yang, C.~Xing, and J.~An, ``Throughput maximization for
  asynchronous {RIS}-aided hybrid powered communication networks,'' \emph{IEEE
  Trans. Wireless Commun.}, vol.~21, no.~6, pp. 4114--4132, 2022.

\bibitem{IRS-U}
H.~Hashida, Y.~Kawamoto, N.~Kato, M.~Iwabuchi, and T.~Murakami,
  ``Mobility-aware user association strategy for {IRS}-aided mm-{W}ave
  multibeam transmission towards {6G},'' \emph{IEEE Journal on Selected Areas
  in Communications}, vol.~40, no.~5, pp. 1667--1678, 2022.

\bibitem{Association-IRS-mmWave}
D.~Zhao, H.~Lu, Y.~Wang, and H.~Sun, ``Joint passive beamforming and user
  association optimization for {IRS}-assisted mmwave systems,'' in \emph{2020
  GLOBECOM}, 2020, pp. 1--6.

\bibitem{S-V-model}
P.~Wang, J.~Fang, X.~Yuan, Z.~Chen, and H.~Li, ``Intelligent reflecting
  surface-assisted millimeter wave communications: Joint active and passive
  precoding design,'' \emph{IEEE Transactions on Vehicular Technology},
  vol.~69, no.~12, pp. 14\,960--14\,973, 2020.

\bibitem{mmWave-multi-IRS}
Y.~Cao, T.~Lv, and W.~Ni, ``Intelligent reflecting surface aided multi-user
  mmwave communications for coverage enhancement,'' in \emph{2020 IEEE 31st
  Annual International Symposium on Personal, Indoor and Mobile Radio
  Communications}, 2020, pp. 1--6.

\bibitem{Fractional-Programming}
K.~Shen and W.~Yu, ``Fractional programming for communication systems part {I}:
  Power control and beamforming,'' \emph{IEEE Transactions on Signal
  Processing}, vol.~66, no.~10, pp. 2616--2630, 2018.

\bibitem{SINR-SOCP}
H.~Xie, J.~Xu, and Y.-F. Liu, ``Max-min fairness in {IRS}-aided multi-cell
  {MISO} systems via joint transmit and reflective beamforming,'' in \emph{2020
  ICC}, 2020, pp. 1--6.

\end{thebibliography}

\end{document}